\documentstyle[prl,preprint,aps,epsf]{revtex}
\begin{document}

\title{Proof of universality for the absorption of massive scalars by the 
higher-dimensional Reissner-Nordstr\"{o}m black holes}

\author{Eylee Jung\footnote{Email:eylee@kyungnam.ac.kr}, 
SungHoon Kim\footnote{Email:shoon@kyungnam.ac.kr} and
D. K. Park\footnote{Email:dkpark@hep.kyungnam.ac.kr 
}, }
\address{Department of Physics, Kyungnam University, Masan, 631-701, Korea}

\maketitle

\maketitle
\begin{abstract}
Motivated by black hole experiments as a consequence of the TeV-scale gravity
arising from modern brane-world scenarios, 
we study the absorption problem for the
massive scalars when the spacetime background is a $(4+n)$-dimensional 
Reissner-Nordstr\"{o}m black hole. For analytic computation we adopt the
near-extreme condition in the spacetime background. 
It is shown that the low-energy
absorption cross section for the s-wave case holds an universality,
{\it i.e.} the absorption cross section equals to the 
area of the black hole horizon divided by  a velocity
parameter.
\end{abstract}

\newpage
\section{Introduction}
In the brane-world scenarios our four-dimensional universe is embedded in the 
higher-dimensional bulk space. Although the higher-dimensional theories have
their own long history\cite{lee84}, modern brane-world scenarios generally
assume the large\cite{ark98-1,anto98} or warped\cite{rs99-1,rs99-2} extra 
dimensions. The most remarkable consequences arising from the modern 
brane-world scenarios seem to be a short-range behavior of the
gravity deviated from the usual Newton 
law\cite{rs99-2,garr00,gidd00,duff00,dvali00,dvali01,jung03-1,park04} and the 
emergence of the Tev-scale gravity. Especially, the emergence of the 
Tev-scale gravity provides a motivation\cite{gidd02-1,dimo01-1,eard02-1} for
the black hole experiments in the future accelerator such as the CERN Large
Hadron Collider. Thus, it is important to investigate the effect of the
extra dimensions in the various properties of black holes. 

In this context, recently, the absorption and emission problems for the 
$(4+n)$-dimensional Schwarzschild black hole are 
examined\cite{harris03-1,jung04-1}. Especially, Ref.\cite{jung04-1}
shows that the ratio factor $1/8$\cite{unruh76} of the low-energy
absorption cross sections for the massive scalar and Dirac fermion minimally
coupled to the four-dimensional Schwarzschild black hole is changed into
$2^{-(n+3) / (n+1)}$ when we have $n$-dimensional toroidally compactified
extra space. 

In this letter we would like to examine the absorption problem of the 
massive scalar minimally coupled to the $(4+n)$-dimensional 
Reissner-Nordstr\"{o}m(RN) black hole. In particular we will show that 
the low-energy absorption cross section for the s-wave massive scalar 
holds the generalized universality\cite{jung04-1}, {\it i.e.}
$\sigma \sim {\cal A}_h / \bar{v}$, where ${\cal A}_h$ is an area 
of the horizon hypersurface and $\bar{v} = \sqrt{1 - m^2/\omega^2}$ is a
velocity parameter introduced firstly in Ref.\cite{unruh76}. In fact, this 
formula is a generalization of the universality for the massless scalar 
examined in Ref.\cite{das97}.

\section{Soultions}
The $(4 + n)$-dimensional RN spacetime is well-known\cite{tang63,myers86} 
whose metric is explicitly 
\begin{equation}
\label{space1}
ds^2 = -\left[1 - \left(\frac{\tilde{r}_+}{\tilde{r}}\right)^{n+1} \right]
        \left[1 - \left(\frac{\tilde{r}_-}{\tilde{r}}\right)^{n+1} \right] dt^2
      + \frac{d \tilde{r}^2}
             {\left[1 - \left(\frac{\tilde{r}_+}{\tilde{r}}\right)^{n+1} \right]
        \left[1 - \left(\frac{\tilde{r}_-}{\tilde{r}}\right)^{n+1} \right]}
      + \tilde{r}^2 d \Omega_{n+2}^2
\end{equation}
where $\tilde{r}_{\pm}$ represent the outer and inner horizons.
Thus, $\tilde{r}_- = 0$ and $\tilde{r}_- = \tilde{r}_+$ means the 
Schwarzschild and extremal limits, respectively. The mass $M$ and charge $Q$
of the RN black hole are given by
\begin{eqnarray}
\label{masscharge}
M&=&\frac{n+2}{16 \pi G} \Omega_{n+2}
   \left(\tilde{r}_+^{n+1} + \tilde{r}_-^{n+1} \right)
                                                      \\   \nonumber
Q&=&\pm \left( \tilde{r}_+ \tilde{r}_- \right)^{\frac{n+1}{2}}
     \sqrt{\frac{(n+1) (n+2)}{8 \pi G}}
\end{eqnarray}
where $\Omega_{n+2}$ is the area of a unit $(n+2)$-sphere
\begin{equation}
\label{hyper}
\Omega_{n+2} = \frac{2 \pi^{\frac{n+3}{2}}}
               {\Gamma \left(\frac{n+3}{2} \right)}.
\end{equation}

For computation of the absorption cross section it is convenient to 
introduce the new parameters $r_0$ and $r_1$ as following:
\begin{equation}
\label{define1}
\tilde{r}_+ = \left(r_1^{n+1} + r_0^{n+1} \right)^{\frac{1}{n+1}}
\hspace{2.0cm}
\tilde{r}_- = r_1.
\end{equation}
Then the spacetime metric (\ref{space1}) reduces to 
\begin{equation}
\label{space2}
ds^2 = -h(r) f^{-n-1}(r) dt^2 + 
f(r) \left[h^{-1}(r) dr^2 + r^2 d \Omega_{n+2}^2 \right]
\end{equation}
where $r$ is a new radial coordinate defined
\begin{equation}
\label{define2}
\tilde{r}^2 = f(r) r^2
\end{equation}
and 
\begin{equation}
\label{define3}
f(r) = \left[1 + \left(\frac{r_1}{r} \right)^{n+1} \right]^{\frac{2}{n+1}}
\hspace{2.0cm}
h(r) = 1 - \left(\frac{r_0}{r}\right)^{n+1}.
\end{equation}
Of course, $d \Omega_{n+2}^2$ is an angle-dependent part of the metric 
\begin{equation}
\label{angle1}
d \Omega_{n+2}^2 = d \theta_1^2 + \sin^2 \theta_1
\left[d \theta_2^2 + \sin^2 \theta_2 
      \left\{ d \theta_3^2 + \cdots + \sin^2 \theta_n
             \left(d \theta_{n+1}^2 + \sin^2 \theta_{n+1} d \phi^2
                              \right) \cdots \right\} \right].
\end{equation}
If $n=1$, the metric (\ref{space2}) exactly coincides with a five-dimensional
solution of the low-energy action of type-IIB string theory compactified
on a torus explicitly introduced in Ref.\cite{hawk97} 
on condition that three charges $Q_1$, $Q_5$ and 
$Q_K$ in Ref.\cite{hawk97} are all same. In this letter, however, we will
consider the arbitrary number of extra dimensions. 
Also we omit the configuration
of the electromagnetic field, which is not necessary throughout the paper.

Now, we consider a scalar field $\Phi$ minimally coupled to the spacetime
(\ref{space2}). Then the usual wave equation $(\Box - m^2) \Phi = 0$
reduces to the following radial equation:
\begin{equation}
\label{radial1}
\left[\frac{h}{r^{n+2}} \partial_r \left(r^{n+2} h \partial_r \right)
- \frac{\ell (\ell + n + 1)}{r^2} h - m^2 f h + f^{n+2} \omega^2 \right]
R = 0
\end{equation}
where we assume the separability 
$\Phi = e^{-i \omega t} R(r) \tilde{Y}_{\ell} (\Omega)$ and 
$\tilde{Y}_{\ell}$ is an higher-dimensional spherical harmonics.
What we want to do is to solve Eq.(\ref{radial1}) in the asymptotic and 
near-horizon regimes separately. Matching one solution with another, we 
would like to extract information on the greybody factor and the absorption 
cross section. 

In the asymptotic region $r \sim \infty$, we have $h \sim f \sim 1$ which 
makes Eq.(\ref{radial1}) simply
\begin{equation}
\label{radial2}
\left[\partial_r^2 + \frac{n+2}{r} \partial_r + \omega^2 \bar{v}^2 \right]
R_{FF}^s = 0
\end{equation}
where $\bar{v} = 1 - m^2 / \omega^2$ and we confined our attention to the 
$s$-wave, {\it i.e.} $\ell = 0$. 
For the case of non-zero angular momentum, {\it i.e.} $\ell \neq 0$, it 
seems to be very difficult to solve Eq.(\ref{radial1}) even in the asymptotic
region. Also the solution should be non-trivially dependent on the number of 
extra dimensions in this case. Since our main interest is focused on the 
universality of the low-energy absorption cross section, we choose
$\ell = 0$ for simplicity.
The subscript of $R$ in Eq.(\ref{radial2}) stands for ``far field'' 
and the superscript indicates our restriction to the $s$-wave.
The solution of Eq.(\ref{radial2}) is easily expressed in terms of the 
Bessel functions:
\begin{equation}
\label{solution1}
R_{FF}^s = r^{-\frac{n+1}{2}}
\left[A_1 J_{\frac{n+1}{2}} (\omega \bar{v} r ) + 
      A_2 Y_{\frac{n+1}{2}} (\omega \bar{v} r ) \right].
\end{equation}

Next we would like to solve the radial equation (\ref{radial1}) in the 
near-horizon region $r \sim r_0$, which corresponds to 
$\tilde{r} \sim \tilde{r}_+$. It is convenient for further analysis to 
introduce a new variable $v = (r_0 / r)^{n+1}$, which changes 
Eq.(\ref{radial1}) into
\begin{equation}
\label{radial3}
\left[(1 - v) \frac{d}{d v} (1 - v) \frac{d}{d v} - 
     \frac{\ell (\ell + n + 1) (1 - v)}{(n+1)^2 v^2} - 
     \frac{m^2 (1 - v) f}{(n+1)^2 r_0^{n+1} v} r^{n+3} + 
     \frac{\omega^2 f^{n+2}}{(n+1)^2 r_0^{n+1} v} r^{n+3} \right] R = 0.
\end{equation}

It seems to be formidable to solve Eq.(\ref{radial3}) in the near-horizon
region $v \sim 1$. Thus, we restrict our attention to the near-extremal
case, {\it i.e.} $r_0 \sim 0$, for analytical analysis. With this 
restriction $f(r)$ can be expanded as a power series with an expansion
parameter $r_0 / r_1$. Then
the radial equation (\ref{radial3}) reduces to 
\begin{equation}
\label{radial4}
\left[(1 - v) \frac{d}{d v} (1 - v) \frac{d}{d v} + 
      \left(D + \frac{C}{v} + \frac{G}{v^2} \right) \right] R^{NE} = 0
\end{equation}
where the superscript $NE$ indicates the near-extremal case and
\begin{eqnarray}
\label{coefficient1}
D&=&\frac{\omega^2 r_0^2}{(n+1)^2}
    \left(\frac{r_1}{r_0} \right)^{2n + 4}
                                                 \\  \nonumber
C&=&\frac{m^2 r_1^2}{(n+1)^2}
    + \frac{2 (n+2) \omega^2 r_0^2}{(n+1)^3} \left(\frac{r_1}{r_0}\right)^{n+3}
    + \frac{\ell (\ell + n + 1)}{(n+1)^2}
                                                 \\   \nonumber
G&=&-\frac{m^2 r_1^2}{(n+1)^2}
     \left[1 - \frac{2}{n+1} \left(\frac{r_0}{r_1} \right)^{n+1} \right]
   + \frac{(n+2) (n+3) \omega^2 r_1^2}{(n+1)^4} - 
     \frac{\ell (\ell + n + 1)}{(n+1)^2}.
\end{eqnarray}
 
In order to solve Eq.(\ref{radial4}) in the near-horizon region, we introduce
$z = 1 - v$ and $R^{NE} = z^{-i (a + b) / 2} F^{NE}$, where the constants
$a$ and $b$ will be fixed later. Then Eq.(\ref{radial4})
becomes
\begin{eqnarray}
\label{radial5}
& & \hspace{1.0cm} z(1 - z) \frac{d^2 F^{NE}}{d z^2} + (1 - i a - i b)(1 - z) 
\frac{d F^{NE}}{d z}
                           \\  \nonumber
&+& \left[ \left( \frac{(a + b)^2}{4} - D \right) + \frac{1}{z}
        \left(D + C + G - \frac{(a + b)^2}{4} \right) + 
        \frac{G}{1 - z}  \right] F^{NE} = 0.
\end{eqnarray}

Since $z \sim 0$ in the near-horizon region, we can simplify Eq.(\ref{radial5})
in this region as 
\begin{eqnarray}
\label{radial6}
& & \hspace{1.0cm} z(1 - z) \frac{d^2 F_{NH}^{NE}}{d z^2} 
+ (1 - i a - i b)(1 - z)
\frac{d F_{NH}^{NE}}{d z}
                           \\  \nonumber
&+& \left[ \left( \frac{(a + b)^2}{4} - D + G \right) + \frac{1}{z}
        \left(D + C + G - \frac{(a + b)^2}{4} \right)  
         \right] F_{NH}^{NE} = 0
\end{eqnarray}
where the subscript $NH$ denotes the near-horizon. Choosing
\begin{eqnarray}
\label{choice1}
a&=& \sqrt{D + C + G} + \sqrt{D - G}
                                        \\   \nonumber
b&=& \sqrt{D + C + G} - \sqrt{D - G},
\end{eqnarray}
we can solve Eq.(\ref{radial6}) in terms of the hypergeometric
functions. Thus, $R^{NE}$ in the near-horizon region, {\it say}
$R_{NH}^{NE}$, becomes
\begin{equation}
\label{solution2}
R_{NH}^{NE} = \alpha_{I} z^{-i \frac{a + b}{2}} 
F\left(-i a, -i b; 1 - ia - ib; z\right) + 
\alpha_{II} z^{i \frac{a + b}{2}} 
F\left(i a, i b; 1 + ia + ib; z \right).
\end{equation}
Since the scalar wave should be purely ingoing in this region due to the
physically relevant reason, we should choose $\alpha_{II} = 0$, which 
yields
\begin{equation}
\label{solution3}
R_{NH}^{NE} = \alpha_{I} z^{-i \frac{a + b}{2}} 
F\left(-i a, -i b; 1 - ia - ib; z\right).
\end{equation}
If we take $z \rightarrow 0$ limit in $R_{NH}^{NE}$, it is easy to show
\begin{equation}
\label{solution4}
R_{NH}^{NE} \sim \alpha_{I} z^{-i \frac{a + b}{2}}
= \alpha_I e^{-i \sqrt{D + C + G} \ln (1 - v)} = \alpha_I e^{-\frac{i}{2}
(a + b) \ln (1 - v)}
\end{equation}
as expected.

\section{Computation of Absorption quantities}
Now, we would like to compute the quantities related to the absorption via
the matching between the near-horizon solution 
(\ref{solution4}) and the asymptotic solution (\ref{solution1}). If we take 
$z \rightarrow 1$ limit in Eq.(\ref{solution3}), it is easy to show
\begin{eqnarray}
\label{matching1}
\lim_{z \rightarrow 1} R_{NH}^{NE}&=& \alpha_{I} z^{-i \frac{a + b}{2}}
\Bigg[ \frac{\Gamma \left(1 - ia - ib\right)}
            {\Gamma (1 - i a) \Gamma (1 - i b) }
       F(-ia, -ib; \epsilon; v)
                                           \\   \nonumber
& &  \hspace{2.0cm}
+ v \frac{\Gamma (1 - ia - ib) \Gamma(-1 + \epsilon)}
         {\Gamma (-ia)  \Gamma (-i b)}
       F(1 - i a, 1 - i b; 2; v)  \Bigg]
\end{eqnarray}
where the regularization parameter $\epsilon = 0^+$ is introduced. Since 
$v \sim 0$ when $z \sim 1$, one can expand Eq.(\ref{matching1}) as a 
power series of $v$. Then the contribution to the linear term 
from the first term in Eq.(\ref{matching1}) involves an infinity
proportional to $1 / \epsilon$. This factor, however, is exactly cancelled out
by the contribution from the second term in Eq.(\ref{matching1}). 
Thus, expanding Eq.(\ref{matching1}) simply yields
\begin{equation}
\label{expand1}
\lim_{r \rightarrow \infty} R_{NH}^{NE} = 
\alpha_I \left[ {\cal E} + {\cal H} \left(\frac{r_0}{r}\right)^{n+1} \right]
\end{equation}
where
\begin{equation}
\label{expand2}
{\cal E} = \frac{\Gamma (1 - ia - ib)}
                {\Gamma (1 -ia)  \Gamma (1 - ib)}
\end{equation}
and ${\cal H}$ is a constant whose explicit expression is not needed. 

Next, we take $r \rightarrow 0$ limit in Eq.(\ref{solution1}), which
is easily carried out by making use of the series expansions of the 
Bessel functions:
\begin{equation}
\label{expand3}
\lim_{r \rightarrow 0} R_{FF}^s = 
A_1 \left[\frac{\left(\frac{\omega \bar{v}}{2}\right)^{\frac{n+1}{2}}}
               {\Gamma \left(\frac{n+3}{2}\right)} + \cdots \right]
+ A_2 \left[-\frac{\Gamma \left(\frac{n+1}{2}\right)
                   \left(\frac{2}{\omega \bar{v}} \right)^{\frac{n+1}{2}}}
                  {\pi}
              \frac{1}{r^{n+1}} + \cdots \right].
\end{equation}

Comparing Eq.(\ref{expand1}) with Eq.(\ref{expand3}) straightforwardly leads
\begin{eqnarray}
\label{matchingfinal}
A_1&=&\alpha_{I} \frac{\Gamma \left( \frac{n+3}{2} \right)}
                      {\left(\frac{\omega \bar{v}}{2}\right)^{\frac{n+1}{2}}}
                  {\cal E}
                                         \\   \nonumber
A_2&=&-\pi \alpha_{I} \frac{r_0^{n+1}}
                {\Gamma \left(\frac{n+1}{2}\right) \left(\frac{2}{\omega\bar{v}
                                                      }\right)^{\frac{n+1}{2}}}
                    {\cal H}.
\end{eqnarray}
It is worthwhile noting $A_1 >> A_2$ in the low-energy approximation,
{\it i.e.} $\omega << 1$, which will be used later. 

In order to compute the greybody factor we take $r \rightarrow \infty$ limit
in $R_{FF}^s$:
\begin{equation}
\label{grey1}
\lim_{r \rightarrow \infty} R_{FF}^s = \varphi_{in} + \varphi_{re}
\end{equation}
where $\varphi_{in}$ and $\varphi_{re}$ are respectively the incident and 
reflected waves, 
whose explicit forms are 
\begin{eqnarray}
\label{grey2}
\varphi_{in}&=&\frac{A_1 + i A_2}
                    {\sqrt{2\pi \omega \bar{v} r^{n+2}}}
               e^{-i \left[\omega \bar{v} r - \frac{n+2}{4} \pi \right]}
                                                          \\  \nonumber
\varphi_{re}&=& \frac{A_1 - i A_2}
                    {\sqrt{2\pi \omega \bar{v} r^{n+2}}}
               e^{i \left[\omega \bar{v} r - \frac{n+2}{4} \pi \right]}.
\end{eqnarray}

From a definition of the conserved flux
\begin{equation}
\label{flux1}
j \equiv \frac{1}{2 i}
\left( \Psi^{\ast} h r^{n+2} \frac{d \Psi}{d r} - \mbox{c.c.} \right),
\end{equation}
the ingoing flux is exactly calculated:
\begin{equation}
\label{flux2}
j_{in} = \frac{1}{2 i}
\left( \phi_{in}^{\ast} h r^{n+2} \frac{d \phi_{in}}{dr} - \mbox{c.c.} \right)
= - \frac{1}{2 \pi} |A_1 + i A_2|^2.
\end{equation}
By the same way it is straightforward to compute the transmitted flux using 
Eq.(\ref{solution4}):
\begin{equation}
\label{flux3}
j_{tr} = - \frac{n+1}{2} (a + b) r_0^{n+1} |\alpha_{I}|^2.
\end{equation}

Thus the greybody factor ${\cal F}$ becomes
\begin{equation}
\label{grey3}
{\cal F} \equiv \Bigg|\frac{j_{tr}}{j_{in}}\Bigg| = (n+1) \pi (a+b) r_0^{n+1}
\Bigg| \frac{\alpha_{I}}{A_1 + A_2} \Bigg|^2
\approx (n+1) \pi (a+b) r_0^{n+1} \Bigg| \frac{\alpha_{I}}{A_1}\Bigg|^2,
\end{equation}
where the last approximation comes from $|A_1| >> |A_2|$ in the low-energy 
approximation. Combining Eq.(\ref{matchingfinal}) and (\ref{grey3}), therefore,
yields
\begin{equation}
\label{grey4}
{\cal F} = (n+1) \pi (a+b) r_0^{n+1}
\frac{\left(\frac{\omega \bar{v}}{2} \right)^{n+1}}
     {\Gamma^2 \left(\frac{n+3}{2} \right)} 
\Bigg| \frac{\Gamma (1 - ia) \Gamma (1 - ib)}{\Gamma (1 - ia - ib)} \Bigg|^2.
\end{equation} 

Using a property of the gamma function 
$|\Gamma (1 - ix)|^2 = \pi x / \sinh x$, it is straightforward to re-express
Eq.(\ref{grey4}) in the form
\begin{equation}
\label{grey5}
{\cal F} = 2 (n+1) \pi^2 a b 
\frac{\left(\frac{\omega \bar{v} r_0}{2} \right)^{n+1}}
     {\Gamma^2 \left(\frac{n+3}{2} \right)}
\frac{e^{2\pi (a+b)} - 1}{(e^{2\pi a} - 1) (e^{2 \pi b} - 1)}.
\end{equation}

Since $D >> C >> G$ which is easily deduced from Eq.(\ref{coefficient1}) 
with a condition of the 
near-extremal limit, Eq.(\ref{choice1}) indicates
\begin{eqnarray}
\label{choice2}
a&\sim& 2 \sqrt{D} = 2 \frac{\omega r_0}{n+1}
\left( \frac{r_1}{r_0} \right)^{n+2} \propto r_0^{-n-1}
                                                      \\   \nonumber
b&\sim& \frac{C}{2 \sqrt{D}} \sim \frac{(n+2) \omega r_1}{(n+1)^2}
\propto r_0^0
\end{eqnarray}
where we fix $\ell = 0$ in Eq.(\ref{coefficient1}) for the restriction 
to the $s$-wave. Eq.(\ref{choice2})
also indicates $a >> b$. With this approximation, therefore, 
the greybody factor
(\ref{grey5}) reduces to 
\begin{equation}
\label{grey6}
{\cal F} \approx 2 \pi r_1^{n+2} \omega
\frac{\left(\frac{\omega \bar{v}}{2} \right)^{n+1}}
     {\Gamma^2 \left(\frac{n+3}{2} \right)}.
\end{equation}

The relation between the absorption cross section and the greybody factor 
in the higher dimensions can
be derived using a $(4+n)$-dimensional optical theorem\cite{gub97}
\begin{equation}
\label{addum1}
\sigma_{\ell}(\omega) = 
\frac{2^n \Gamma \left( \frac{n+3}{2} \right)}
     {\pi (\omega \bar{v} r_H)^{n+2}}  {\cal A}_h
\frac{(2 \ell + n + 1) (\ell + n)!}
     {(n+1)! \ell !} {\cal F}
\end{equation}
where $r_H$ and ${\cal A}_h$ are respectively the horizon radius and 
the area of the
horizon hypersurface. 
For the s-wave case Eq.(\ref{addum1}) implies   
\begin{equation}
\label{section1}
\sigma_s(\omega) = \frac{2^n \Gamma^2 \left(\frac{n+3}{2}\right)}
                        {\pi (\omega \bar{v} r_1)^{n+2}}
                   {\cal A}_h {\cal F} = 
\frac{{\cal A}_h }{\bar{v}}
\end{equation}
where ${\cal A}_h$ is for our case
\begin{equation}
\label{hyper-surface}
{\cal A}_h = \frac{2 \pi^{\frac{n+3}{2}}}
                  {\Gamma \left(\frac{n+3}{2} \right)}
             r_1^{n+2}.
\end{equation}
Thus the universality for the low-energy absorption cross section is valid 
in the  higher-dimensional RN as well as Schwarzschild black holes.

\section{Conclusion}
In this letter we have examined the absorption problem for the massive 
scalars by the higher-dimensional RN black holes. For analytic computation 
we have chosen the near-extreme condition of the RN black holes. However, it 
seems to be greatly nice if we can remove this near-extreme condition. If it
is possible, we may be able to study the absorption and emission problems in
the Schwarzschild and RN black holes as an unified way. Apparently, it is 
beyond the dilute gas region\cite{mal97-1}. To relax the near-extremal 
condition we should solve Eq.(\ref{radial3}) in the near-horizon region without
relying on $r_0 \sim 0$, which seems to be an highly non-trivial problem. 
Although it is assumed to be possible to solve it approximately, 
we should check whether this solution 
gives physically reasonable results through a matching with the asymptotic 
solution. This should be another difficult problem. 
Perhaps, numerical approach can be a breakthrough. For this we may use the 
numerical method of Ref.\cite{jung04-2} from the beginning. It may be 
possible in this approach to relax the s-wave restriction too. 
We hope to consider this elsewhere.

\vspace{1cm}

{\bf Acknowledgement}:  
This work was supported by the Korea Research Foundation
Grant (KRF-2003-015-C00109).

\end{document}